\documentclass[doublecol]{epl2}
\usepackage{color}
\usepackage{bm}
\usepackage{latexsym}
\usepackage{amssymb}
\usepackage{amsmath}
\usepackage{graphicx}
\usepackage{ulem}
\begin{document}

\newcommand{\half}{\mbox{\small $\frac{1}{2}$}}
\newcommand{\cc}{\mbox{c.c.}}
\newcommand{\abs}[1]{\lvert #1 \rvert}
\newcommand{\avg}[1]{\langle #1 \rangle}
\newcommand{\bra}[1]{\left\langle #1\right|}
\newcommand{\ket}[1]{\left|#1\right\rangle }
\newcommand{\edit}[1]{\textcolor{red}{#1}}
\newcommand{\nin}{n_{in}}
\renewcommand{\etal}{\textit{et al.}}

\title{The crossover from strong to weak chaos for nonlinear waves in disordered systems} 

\author{T. V. Laptyeva \and J. D. Bodyfelt \and D. O. Krimer \and Ch. Skokos \and S. Flach}
\shortauthor{T. V. Laptyeva \etal}
\institute{Max Planck Institute for the Physics of Complex Systems, N\"othnitzer Stra\ss e 38, D-01187 Dresden, Germany}

\pacs{05.45.-a}{Nonlinear dynamics and Chaos}
\pacs{05.60.Cd}{Classical transport}
\pacs{63.20.Pw}{Phonons in crystal lattices - Localized modes}

\abstract{
We observe a crossover from strong to weak chaos in the spatiotemporal evolution of multiple site excitations within disordered chains with cubic nonlinearity. 
Recent studies have shown that Anderson localization is destroyed, and the wave packet spreading is characterized by an asymptotic divergence of the second moment $m_2$ in time (as $t^{1/3}$), due to weak chaos.
In the present paper, we observe the existence of a qualitatively new dynamical regime of strong chaos, in which the second moment spreads even faster (as $t^{1/2}$), with a crossover to the asymptotic law of weak chaos at larger times. 
We analyze the pecularities of these spreading regimes and perform extensive numerical simulations over large times with ensemble averaging. A technique of local derivatives on logarithmic scales is developed in order to quantitatively visualize the slow crossover processes. 
}

\maketitle

Wave propagation within random potentials is an interdisciplinary research field applicable to many diverse systems;
regardless of their classical or quantum nature, the overall wave behavior provides common ground for understanding 
transport properties. 
One such property - theoretically predicted by Anderson\cite{A58} and since labeled ``Anderson localization'' (AL) - is
a halt of wave propagation due to exponentially localized normal modes (NMs) of the random potential. The
significance of AL has been evidenced in the past decades by a bevy of experimental observation; including 
optics\cite{optic}, acoustics\cite{acoustic}, microwaves\cite{microwave}, and matter waves\cite{matter}.

In many experimental situations, AL can be strongly altered by the appearance of nonlinearity within the potential. 
These nonlinearities can be induced via the nonlinear Kerr effect in disordered photonic lattices\cite{NL_photonic},
or atomic Bose-Einstein condensate interactions (controlled by Feshbach resonances) in optical lattices\cite{NL_BECs}.

The question of the interplay between disorder and nonlinearity - how the two complement, frustrate, or reinforce each other - is thus
of strong importance. The theoretical study of AL in random nonlinear lattices has been advanced 
using several approaches including the studies of transmission\cite{transmission} and stationary solutions\cite{stationary}.
Recent research in dynamics of wave spreading within nonlinear disordered media focuses on the spatiotemporal evolution 
of wave packets, debating the asymptotic spreading law against an eventual blockage\cite{dynamics,PS08,kopidakis,VKF09,FKS09,SKKF09}. 
Previous numerical studies show that the second moment of a wave packet starting from a single site excitation grows as $t^{1/3}$\cite{FKS09,SKKF09}. However when starting from
a distributed single normal mode state, faster growth is reported, though not quantitatively assessed\cite{FKS09,SKKF09,SF10}. Theoretical expectations range from spreading without limits
to a slowdown and restoration of AL. Different spreading characteristics are also claimed to be $t^{2/5}$\cite{PS08},
$t^{1/3}$\cite{FKS09,SKKF09}, and a two regime case with $t^{1/2}$ and asymptotic $t^{1/3}$\cite{F10}.
This letter aims to clarify some of these controversies. 

We first show that using estimates for average spacings and nonlinear frequency shifts, three different evolution regimes can be identified. 
In particular, in contrast to previous results we obtain a new regime of strong chaos which is accessible by initial multiple site excitations, 
but \textbf{not} by single site excitations. Contrary to previous studies, we perform extensive ensemble averaging over $1000$ disorder realizations.
As a result, smooth functional dependencies of wave packet characteristics on time are obtained. Using a technique of smoothing and local 
differentiation on logarithmic scales, we are able to observe the fast spreading regime of strong chaos, proceeded by the predicted crossover into 
the asymptotic regime of weak chaos, as argued in ref.\cite{F10}. No further saturation and slowing down of the asymptotic spreading process is observed on the longest 
time scales of observation.

We study two different Hamiltonian models. The first is the one-dimensional disordered nonlinear Schr\"{o}dinger (DNLS) 
equation
\begin{equation}
\mathcal{H}_D = \sum_l \epsilon_l \abs{\psi_l}^2 + \frac{\beta}{2} \abs{\psi_l}^4 - (\psi_{l+1} \psi_l^\star + \cc) \label{eq:DNLS}
\end{equation}
in which $\epsilon_l$ is the onsite energy chosen uniformly from a $[-W/2, W/2]$ random distribution. 
By $\partial_t \psi_l = \partial \mathcal{H}_D / \partial (i \psi_l^\ast)$, the equations of motion are generated:
\begin{equation}
i \dot{\psi}_l = \epsilon_l \psi_l + \beta \abs{\psi_l}^2 \psi_l - \psi_{l+1} - \psi_{l-1} \;. \label{eq:DNLSEOM}
\end{equation}
The second model considered is oscillators on a quartic Klein-Gordon (KG) lattice, given as
\begin{equation}
\mathcal{H}_K = \sum_l \frac{p_l^2}{2}+\frac{{\tilde \epsilon}_l}{2} u_l^2 + \frac{1}{4}u_l^4 + \frac{1}{2W}(u_{l+1}-u_l)^2 \label{eq:KG}
\end{equation}
where $u_l$ and $p_l$ are respectively the generalized coordinate/momentum on the site $l$ with an energy of $\mathcal{E}_l$, 
and $\tilde{\epsilon_l}$ are the disordered potential strengths chosen uniformly in $[1/2,3/2]$. 
Likewise, $\partial_t^2 u_l = - \partial \mathcal{H}_K / \partial u_l$ generates the equations of motion 
\begin{equation}
\ddot{u}_l = -{\tilde \epsilon}_l u_l - u_l^3 + \frac{1}{W}(u_{l+1}+u_{l-1} - 2u_l) \;. \label{eq:KGEOM}
\end{equation}
The two models have only onsite cubic nonlinearity, but the specific methods discussed here can also be generally applied to other nonlinear models
with other powers of nonlinearity or long range nonlinear dependencies.

Similar to $\beta$ in the DNLS case, the nonlinear control parameter for the KG model is the total energy, $E = \sum_l \mathcal{E}_l \geq 0$.
Both models conserve the total energy; the DNLS model also conserves the total norm ${\cal S} = \sum_{l} \abs{\psi_l}^2$. 
As a practical note between the two models: the KG model is much more computationally friendly, with numerical integration speeds of 
two orders of magnitude faster. For small amplitudes an approximate mapping, $\beta {\cal S} \approx 3WE$, from the KG model to the 
DNLS model exists\cite{smallapprox}. Further analytics will, in general, be discussed in terms of the DNLS model, since it now 
straightforward to adapt similar results for the KG model using the aforementioned mapping.

By neglecting the nonlinear terms, the DNLS model (\ref{eq:DNLS}) reduces to the linear eigenvalue problem $\lambda A_l = \epsilon_l A_{l} - (A_{l+1} + A_{l-1})$. 
This leads to a set of NM amplitudes, $A_{\nu,l}$, with NM frequencies of $\lambda_{\nu} \in [-W/2-2, W/2 + 2]$ in a spectrum width of $\Delta = 4 + W$.
The coefficient choice of $1/(2W)$ in the KG model (\ref{eq:KG}) allows a linear reduction to the same eigenvalue problem as for the DNLS, but with the values of
$\epsilon_l = W(\tilde{\epsilon}_l - 1)$ and $\lambda_\nu = W\omega_\nu^2 - W - 2$; in this case, $\Delta=1+4/W$ is the width of the squared eigenfrequency 
spectrum, $\omega_{\nu}^2 \in [1/2, 3/2 + 4/W]$. 

The NM asymptotic spatial decay is given by $A_{\nu, l} \sim e^{-l/\xi(\lambda_\nu)}$ where $\xi(\lambda_\nu) $ is the localization length. 
It is approximated\cite{KM93} in the limit of weak disorder ($W \ll 1$) as $\xi{(\lambda_\nu)} \le \xi(0) \approx 96 W^{-2}$. 
The NM participation number $P_{\nu} = 1/\sum_{l}A^4_{\nu,l}$ characterizes the NM spatial extent. 
An average measure of this extent is the localization volume $V$, which is on the order of $3.3 \xi(0)$ for weak disorder
and unity in the limit of strong disorder\cite{KF10}. The average frequency spacing of NMs within a localization volume is then $d \approx \Delta/V$. The two frequency 
scales $d < \Delta$ are thus expected to determine the packet evolution details in the presence of nonlinearity. 

Nonlinearity induces an interaction between NMs. The variables $\phi_\nu = \sum_l A_{\nu,l} \psi_l$ determine the complex time-dependent 
amplitudes of the NMs. Since all NMs are exponentially localized in space, each of them is effectively coupled to a finite number of neighbor 
modes, \textit{i.e.} the interaction range is finite. However, the strength of this coupling is proportional to a characteristic norm density,
$n = |\phi|^2$. The frequency shift due to the nonlinearity is then $\delta \sim \beta n$.

We track the normalized forms of the NM norm densities, $z_\nu \equiv \abs{\phi_\nu}^2 / \sum_{\mu} \abs{\phi_\mu}^2$, for DNLS.
The KG counterpart is normalized energy density distributions. These densities are sorted on the center-of-norm coordinate
$X_\nu = \sum_l l A_{l,\nu}^2$, and two measures in NM space used: the participation number $P = 1/\sum_\nu z_\nu^2$ which queries the 
quantity of strongest excited sites, and the second moment $m_2 = \sum_\nu (\nu - \bar \nu)^2 z_\nu$ (where $\bar \nu = \sum_\nu z_\nu$) which probes 
distances between a distribution's tail and center. The ratio of the two measures $\zeta = P^2 / m_2$ (the compactness index\cite{SKKF09}) 
quantifies the sparsity of a packet - thermalized distributions have $\zeta \approx 3$, while $\zeta \ll 3$ indicates either very sparse packets, 
or partial self-trapping. 

We consider compact wave packets at $t=0$ spanning a width $L$ centered in the lattice, such that within $L$ there is a constant 
initial norm density of $\nin$ and a random phase at each site (outside the volume $L$ the norm density is zero). 
In the KG case, this equates to exciting each site in the width $L$ with the same energy density, $\mathcal{E}=E/L$, 
\textit{i.e.} initial momenta of $p_l = \pm \sqrt{2\mathcal{E}}$ with randomly assigned signs. 

For $\beta=0$ and  $L < V$, the packet will extend over $V$ during the time $\tau\sim 2\pi/d$; after that, AL stops further spreading, 
with a lowered norm density $n(\tau) \approx (\nin L)/V$.
For $L \geq V$, the norm density will not change appreciably up to $\tau$, so $n(\tau) \approx \nin$. 
For $\beta > 0$, the nonlinear frequency shift should be compared with the average spacing $d$.
If $\beta n(\tau) < d$, most of the NMs are weakly interacting with each other; hence, this regime is dubbed ``weak chaos''.
Once $\delta=d$, the NM frequency renormalization begins to allow some NMs to interact resonantly, \textit{i.e.} strongly, with each other.
If $\beta n(\tau)  > d$, almost all NMs in the packet are resonantly interacting. This regime will be coined ``strong chaos''.
If $\delta \geq \Delta $, a substantial part of the wave packet will be self-trapped.
This is due to the occurrence of a wave packet frequency shifting out of resonance with the finite-width linear spectrum.
In fact, partial self-trapping will occur already for $\delta \geq 2$ since at least some sites in the packet may be tuned out of
resonance. For a single-site excitation $L = 1$ the strong chaos regime shrinks to zero width and one is left only with either weak chaos 
or self-trapping\cite{PS08,FKS09,SKKF09,VKF09}. 
\textbf{The key distinction of the multiple-site excitations is that they can spread in the strong chaos regime.}

\begin{figure}[htb]
\includegraphics[width=\columnwidth,keepaspectratio,clip]{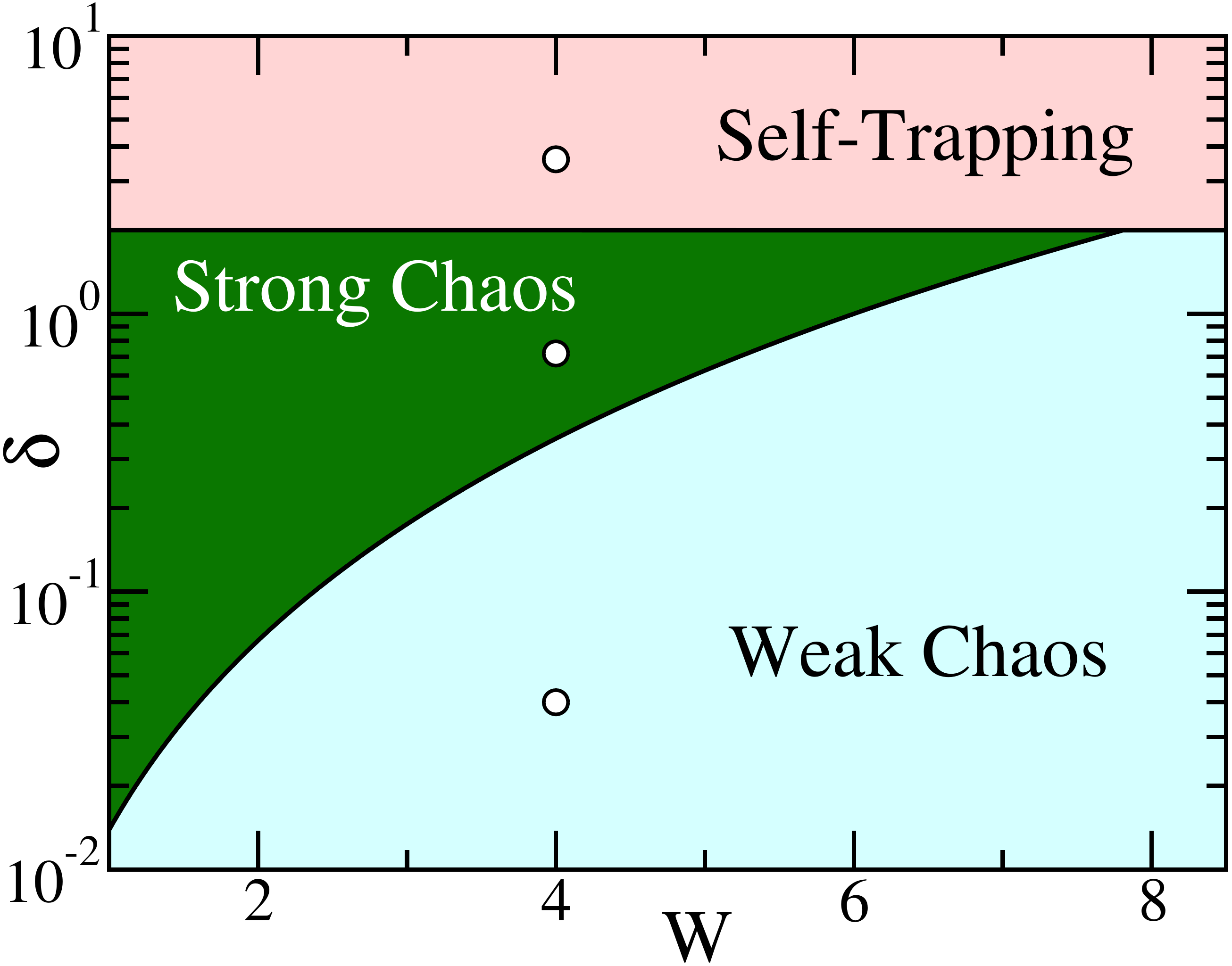}
\caption{(Color online) Parametric space of disorder, $W$, vs. the frequency shift induced by nonlinearity, $\delta$, for the DNLS model. 
The KG analog is obtained by the small amplitude mapping $\mathcal{E} \rightarrow 3W\delta$ (see inset of fig.\ref{fig:3}). 
Three spreading regimes are shown for dynamics dictated by: (i) weak chaos (pale blue), (ii) strong chaos (green), and 
(iii) the onset of self-trapping (pale red). 
The three circles show the initial numerical values used in fig.\ref{fig:2}.}
\label{fig:1}
\end{figure}

Fig.\ref{fig:1} summarizes the predicted regimes in a parametric space for the case $L = V$, in which lines represent 
the regime boundaries $ \delta = d$ and $\delta = 2$. The lower boundary is analytically found, via $d=\Delta/(3.3 \xi(0))$ 
with $\xi(0)=96W^{-2}$ being the weak disorder estimate. More sophisticated numerical estimates of $d$\cite{KF10} yield only 
slight corrections for $W > 6$. It should be noted that the regime boundaries in fig.\ref{fig:1} are NOT sharp, rather there 
is some transitional width between the regimes. The weaker the strength of disorder, the larger the window of strong chaos. Inversely, for 
$W \geq 8$ the strong chaos window closes almost completely. Ideally, we should utilize the smallest possible value of $W$.
Computational limits restrict this, so we choose a reference of $W=4$. 
It is important to note that $\delta$ will be reduced in time, since a spreading wave packet increases in size and drops its 
norm (energy) density. This gives the following interpretation of fig.\ref{fig:1}: given an initial norm density, the packet is in 
one of the three regimes (for example, the three circles in fig.\ref{fig:1}). A packet launched in the weak chaos regime stays in
this regime, while one launched in the strong chaos regime spreads to the point that it eventually crosses over into the asymptotic regime 
of weak chaos at later times.

In order to observe the crossover, we use $L=21$ (which is approximately equal to $V$ for $W=4$) in system sizes of $1000-2000$ sites. 
For DNLS, an initial norm density of $\nin=1$ is used, so that initially $\delta \sim \beta$. Nonlinearities ($\mathcal{E}$ for KG) 
are chosen within the three spreading regimes (see fig.\ref{fig:1}), respectively  $\beta \in \left\lbrace 0.04, 0.72, 3.6\right\rbrace$ and 
$\mathcal{E} \in \left\lbrace 0.01, 0.2, 0.75\right\rbrace$. Eqs.(\ref{eq:DNLSEOM},\ref{eq:KGEOM}) are time evolved using SABA-class 
split-step symplectic integration schemes\cite{SKKF09,LR01}, with time-steps of $dt \sim 10^{-2}-10^{-1}$ up to a maximum $t \sim 10^7-10^9$. 
Energy conservations are accurate to $<0.1\%$ (for discussion of numerical accuracy in symplectic integrators, please see ref.\cite{SKKF09}).

\begin{figure}[htb]
\includegraphics[width=\columnwidth,keepaspectratio,clip]{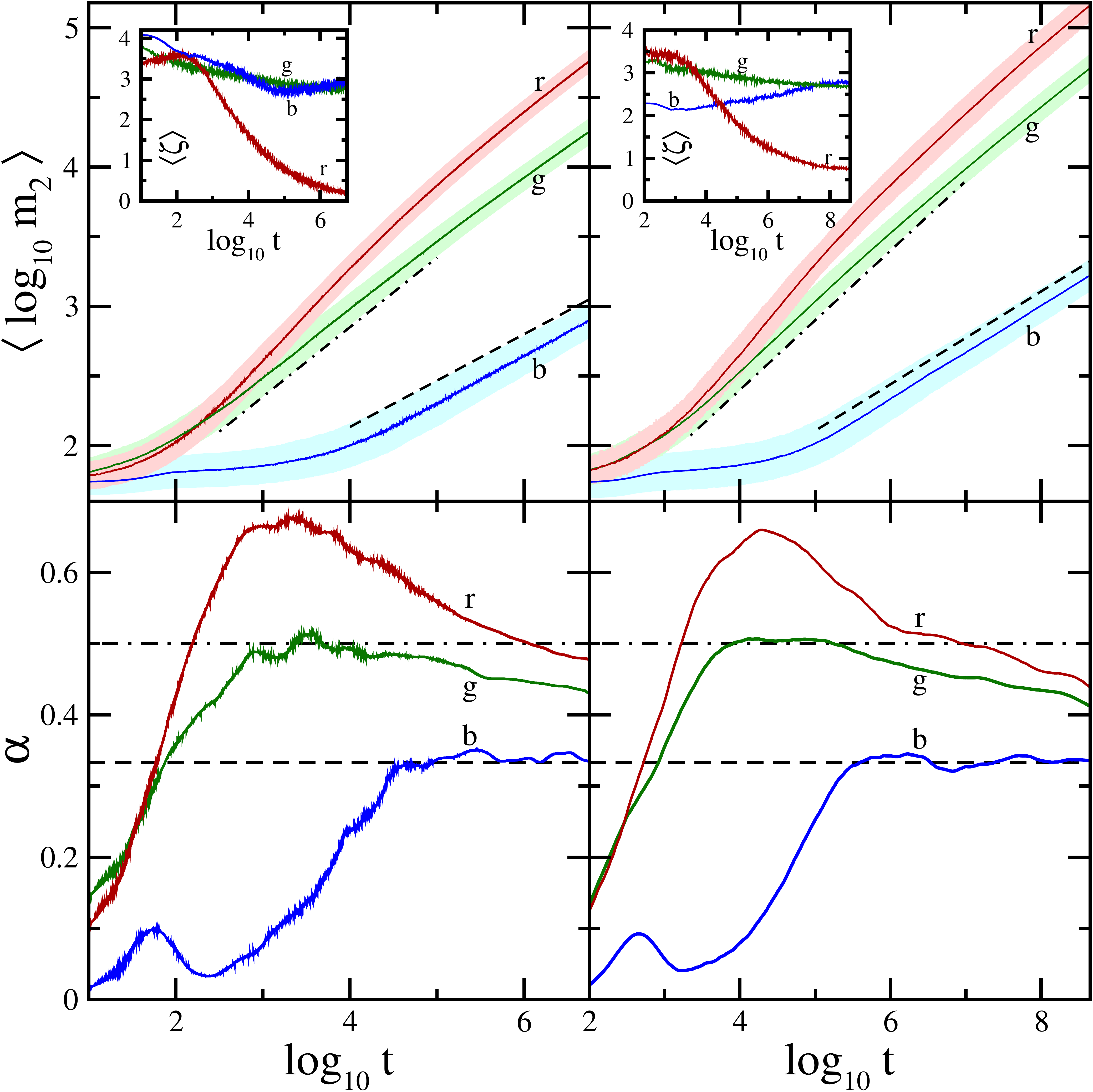}
\caption{
(Color online) Upper row: Average log of second moments (inset: average compactness index) vs. log time for the DNLS (KG) on the 
left (right), for $W=4, L=21$. Colors/letters correspond the three different regimes: 
(i) weak chaos - (b)lue, $\beta =0.04 \, (\mathcal E=0.01)$, 
(ii) strong chaos - (g)reen, $\beta = 0.72 \, (\mathcal E=0.2)$,
(iii) self-trapping - (r)ed, $\beta = 3.6 \, (\mathcal E=0.75)$.
The respective lighter surrounding areas show one standard deviation error. Dashed lines are to guide the eye to $\sim t^{1/3}$, 
while dotted-dashed guides for $\sim t^{1/2}$.
Lower row: Finite difference derivatives for the smoothed $m_2$ data respectively from above curves.
}
\label{fig:2}
\end{figure}
Ensemble averages over disorder were calculated for $1000$ realizations and are shown in fig.\ref{fig:2} (upper row).
In the regime of weak chaos we find a subdiffusive growth of $m_2$ at large times according to $m_2 \sim t^\alpha, \, \alpha \le 1$,
with a compactness index $\zeta \approx 3$. Note that the subdiffusive growth is difficult to see initially in fig.~\ref{fig:2}
for two reasons. Firstly, the logarithmic scaling hides any small initial growth, and secondly, there is a
characteristic time scale for the packet to spread from its initial preparation. 
In the regime of strong chaos we observe a faster subdiffusive growth of $m_2$, with an additional slowing down at larger times, as expected 
from the predicted crossover. The compactness index is also $\zeta \approx 3$, as in the weak chaos regime. Finally, in the regime of partial 
self-trapping $m_2$ grows, but the compactness index $\zeta$ decreases in time substantially. This indicates that a part of the wave packet 
is arrested, and another part is spreading.

In order to quantify our findings, we smooth $\avg{\log m_2}$ with a locally weighted regression algorithm\cite{CD88}, and then apply a central finite difference
to calculate the local derivative 
\begin{equation}
\alpha(\log t) = \frac{{\rm d} \avg{\log m_2}}{ {\rm d} \log t}\;.
\label{alpha(t)}
\end{equation}
The outcome is plotted in the lower row in fig.\ref{fig:2}.

In the weak chaos regime the exponent $\alpha(t)$ increases up to $1/3$ and stays at this value for later times.
In the strong chaos regime $\alpha(t)$ first rises up to $1/2$, keeps this value for one decade, and then drops down, as predicted. Finally, in the self-trapping
regime we observe an even larger rise of $\alpha(t)$. Additionally, we also performed numerics for $W \in \left\lbrace 1,2,6\right\rbrace$ with respective initial 
packetwidths of $L=V \in \left\lbrace 361, 91, 11\right\rbrace $. Results are qualitatively similar to those shown in fig.~\ref{fig:2}, and thus omitted
for graphical clarity.

Following the analysis in ref.\cite{F10}, the modes inside the packet interact in a nonintegrable way leading to chaotic dynamics. The norm (energy) diffusion
is characterized by a diffusion rate $D \sim \beta^2n^2(\mathcal{P}(\beta n))^2$, where $\mathcal{P}(\beta n)\approx 1-e^{-\beta n/d}$ is the probability of packet modes to 
resonate\cite{F10,KF10}. It is this resonance probability that largely dictates whether the chaos is strong or weak.
From $m_2 \sim 1/n^2$ and the diffusion equation $m_2 \sim Dt$, one obtains an equation $1/n^2 \sim \beta(1-e^{-\beta n/d})t^{1/2}$ that determines the subdiffusive spreading 
crossover from the regime of strong chaos to that of weak chaos
\begin{equation}
 m_2 \sim 
\begin{cases}
\beta t^{1/2},  & \beta n/d > 1  \text{ (strong chaos)} \\
d^{-2/3}\beta^{4/3} t^{1/3}, & \beta n/d < 1 \text{ (weak chaos)} \\
\end{cases}\label{eq:WSC}
\end{equation}

The impact of the strong chaos regime is seen in the resonance probability; $\mathcal{P} \approx 1$ if $\beta n$ is sufficiently larger than $d$. Such a situation can be generated for packets with large enough 
$\beta n$ (or energy density $\mathcal{E}$ for KG) in which every mode in the packet resonates, and the condition for strong chaos yields faster spreading, $m_2 \sim t^{1/2}$. 
These predictions for strong chaos are then observed at $t \sim 10^3-10^4$ (KG: $10^4-10^5$) in fig.\ref{fig:2}; time averages in these regions over the green curves yield
$\alpha \approx 0.49\pm 0.01 \;(\mbox{KG: } 0.51\pm 0.02)$. 

With spreading continuing in the strong chaos regime, the norm density in the packet will decrease, and eventually $\beta n \leq d$. Then a dynamical crossover occurs to the slower weak chaos subdiffusive spreading. This crossover spans logarithmic time scales. Nevertheless, in the green curves of fig.\ref{fig:2} clear decay in $\alpha$ to values below $1/2$ is observed. Fits of the decay further suggest $\alpha \approx 1/3$ at $t\sim 10^{10} - 10^{11}$. The challenge remains to directly observe saturation at times accessible in computational experiments. 

\begin{figure}[htb]
\includegraphics[width=\columnwidth,keepaspectratio,clip]{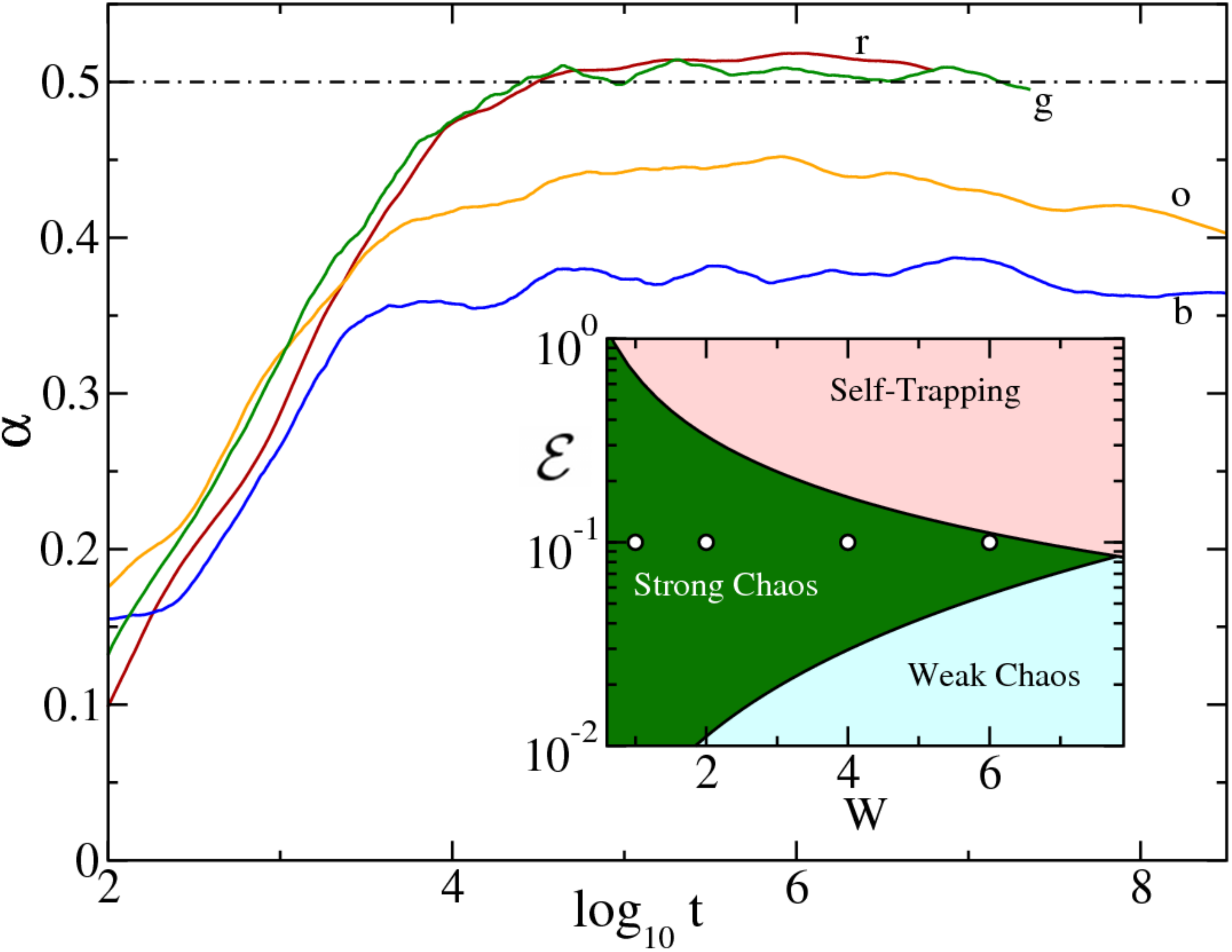}
\caption{(Color online) Spreading behavior in the strong chaos regime for the KG model, with an initial energy density of 
$\mathcal{E}=0.1$. The four curves are for the disorder strengths of: $W=1$ - (r)ed, $W=2$ - (g)reen, $W=4$ - (o)range,
$W=6$ - (b)lue. Inset: the KG analog of the DNLS parametric space, fig.\ref{fig:1}. The four points correspond to the disorder
strengths used in the main portion of the figure.}
\label{fig:3}
\end{figure}

The duration of $\alpha = 1/2$ (and thus when the crossover occurs) is largely dependent on how deep in the strong chaos regime the state is initially. Since the boundaries between different regimes are NOT sharp, but rather have some characteristic width, ideally one should utilize the smallest possible value of $W$. This is shown in fig.\ref{fig:3} for the KG model. For $W \in \left\lbrace 1,2\right\rbrace $, a long plateau at $\alpha = 1/2$ is clearly observed. For $W \in \left\lbrace 4,6 \right\rbrace$, the initial energy density approaches one of the boundary lines and likely crosses into a boundary window, in which $\alpha < 1/2$. 

In the regime of self-trapping, a good portion of the excitation remains highly localized, while the remainder spreads (red curves in fig.\ref{fig:2}). Therefore, $P$ does not grow significantly, but the second moment does. Consequently, $\zeta$ drops and is a good indicator of the degree of self-trapping. The time evolution of $\zeta$ for excitations in different regimes is shown in the insets of fig.\ref{fig:2}. 
In the regimes of weak and strong chaos, if self-trapping is avoided, the compactness index at largest computational times is $\zeta \approx 2.85\pm0.79 \; (\mbox{KG: }2.74\pm0.83)$, as seen in the blue and green curves of fig.\ref{fig:2}. This means that the wave packet spreads, but remains thermalized ($\zeta \approx 3$). For the self-trapping regime (red curves), the compactness index asymptotically decreases to very small values.
Note in fig.\ref{fig:2} at intermediate times, there is transient growth where $\alpha > 1/2$; nonetheless it remains subdiffusive ($\alpha<1$). At larger times, this overshoot decreases. This is presumably due to some self-trapped states which interact strongly with the spreading part of the packet and release their norm (energy) into the thermal cloud at some time. These more complicated scenaria are not yet quantitatively understood,
and certainly remain for future exploration. 

This crossover can be expected to show up in measurements of the heat conductivity $\kappa$ at finite norm (energy) densities. 
According to the heat equation $\partial T(x,t)/\partial t = (\kappa /c) \partial^2 T / \partial x^2$ where $T$ is the temperature and $c$ the specific heat.
Therefore  the heat conductivity is proportional
to the diffusion rate $\kappa = cD$. For small norm (energy) densities, heat is proportional to the densities. 
Therefore we expect that for $\beta n > d$, \textit{i.e.} in the regime of strong chaos, $\kappa \sim T^2$ (here $T$ is the temperature).
For small enough temperatures one crosses over into the regime of weak chaos, and consequently we expect $\kappa \sim T^4$.

Let us summarize. In the presence of nonlinearity within one-dimensional disordered systems, Anderson localization is destroyed. In this Letter, we use a technique
of ensemble averaging and local derivatives on logarithmic scales. In contrast to previous results for single site excitations, 
we find that multiple site excitations can evolve either in the asymptotic regime of weak chaos, or in an intermediate regime of strong chaos (excluding self-trapping
for strong nonlinearities). In the weak chaos regime the second moment $m_2$ grows subdiffusively as $t^{1/3}$. In the strong chaos regime subdiffusion is faster,
yielding $m_2 \sim t^{1/2}$, with a subsequently slow (on logarithmic time scale) crossover to an asymptotic weak chaos law. 

\acknowledgements
The authors wish to thank S.~Aubry, S.~Fishman, N.~Li, R.~Khomeriki, \& A.~Pikovsky for insightful discussions.


\end{document}